# Strain analysis of multiferroic $BiFeO_3$-$CoFe_2O_4$ nanostructures by Raman scattering

O. Chaix-Pluchery[1], C. Cochard[1], P. Jadhav[1], J. Kreisel[1,*]

[1] Laboratoire des Matériaux et du Génie Physique, Grenoble INP, CNRS, Minatec, 3, parvis Louis Néel, 38016 Grenoble, France

N. Dix[2], F. Sánchez[2], J. Fontcuberta[2]

[2] Institut de Ciència de Materials de Barcelona (ICMAB–CSIC), Campus UAB, Bellaterra 08193, Spain

**Abstract**

We report a Raman scattering investigation of columnar $BiFeO_3$-$CoFe_2O_4$ (BFO-CFO) epitaxial thin film nanostructures, where BFO pillars are embedded in a CFO matrix. The feasibility of a strain analysis is illustrated through an investigation of two nanostructures with different BFO-CFO ratios. We show that the CFO matrix presents the same strain state in both nanostructures, while the strain state of the BFO pillars depends on the BFO/CFO ratio with an increasing tensile strain along the out-of-plane direction with decreasing BFO content. Our results demonstrate that Raman scattering allows monitoring strain states in complex 3D multiferroic pillar/matrix composites.

* Corresponding author:   *jens.kreisel@grenoble-inp.fr*

Multiferroic materials which posses simultaneously several so-called ferroic orders such as ferromagnetism, ferroelectricity and/or ferroelasticity, currently attract a considerable interest.[1-3]

To overcome the scarcity of single-phase multiferroics, and to provide new magnetoelectric coupling mechanism at room temperature, recent work concentrates on the class of artificial multiferroics in the form of composite-type materials or thin film nano-/hetero-structures.[4-8] In such systems, it is the elastic coupling interaction between the magnetostrictive phase and the piezoelectric phase that leads to the observed magnetoelectric response. One of the appealing composite-type structures is a self-assembled nanostructure of ferro(i)magnetic nanopillars embedded in a ferroelectric matrix, and vice versa. Since the 2004 work on $BaTiO_3$–$CoFe_2O_4$ nanostructures by Zheng *et al.* [4], several other combinations with immiscible $ABO_3$ perovskites and $AB_2O_4$ spinels have been reported.[5,8,9]

In this study we present an investigation of $BiFeO_3$-$CoFe_2O_4$ (BFO-CFO) nanostructures by Raman scattering, which has shown to be a versatile probe for investigating structural properties in thin oxide films.[10-15] Most Raman scattering studies of composite-type thin films in the literature concern thin film superlattices.[16-19] Only one Raman study of pillar-matrix nanostructures related to the analysis of a $BaTiO_3$–$CoFe_2O_4$ nanostructure is reported; it is restricted to only one band of the CFO spinel pillars[20]. The aim of this study is to demonstrate the feasibility of a more detailed Raman scattering-based analysis of the strain state of both pillars and matrix materials in columnar multiferroic nanostructures.

BFO-CFO nanostructures, 100 nm of thickness, were deposited on $SrTiO_3$ (STO) (111) substrates by pulsed laser deposition (KrF laser, 5 Hz), using a BFO-CFO target with molar ratio of 65:35 (65BFO-35CFO) or 35:65 (35BFO-65CFO)[21]. We have earlier reported structural and microstructural characterization of some of such nanostructures.[8,22] From X-ray diffraction experiments, it is found that the out-of-plane $d_{111}$ spacing of BFO ($d_{111}$(BFO)) of the nano-composite is slightly expanded (2.312(1) Å) compared to BFO reference values ($d_{006}$=2.307 Å, ICSD 01-077-4901, $R3c$ (SG 161)). On the contrary, the $d_{111}$ (CFO) of the CFO fraction of the nano-composite is slightly compressed as compared to an *ad-hoc* prepared 100 nm reference CFO film on STO (111) (4.838(2) Å and 4.857(5) Å, respectively).

Micro-Raman spectra were recorded using a 488.0 nm laser line through a LabRam Jobin-Yvon spectrometer with a spectral cut-off around 100 cm$^{-1}$. Our experiments have been carried out using laser powers of less than 1 mW into a focused spot of 1 μm$^2$ under the microscope to avoid overheating. The reproducibility of spectra on different places of the sample has been verified. The fitting procedure of the Raman lines was performed using individual lorentzian profiles after baseline subtraction.

The top Raman spectrum of Figure 1 presents the spectral signature of a 65BFO-35CFO thin film nanostructure in the range from 100 to 1600 cm$^{-1}$. Four regions can be distinguished: a first region I with three rather sharp and well-separated bands in the low wavenumber range below 250 cm$^{-1}$, a more complex region II of several overlapping bands below 550 cm$^{-1}$, a region III between 550 and 900 cm$^{-1}$ with two dominating bands, and a region IV between 900 and 1400 cm$^{-1}$ with broad overlapping bands.

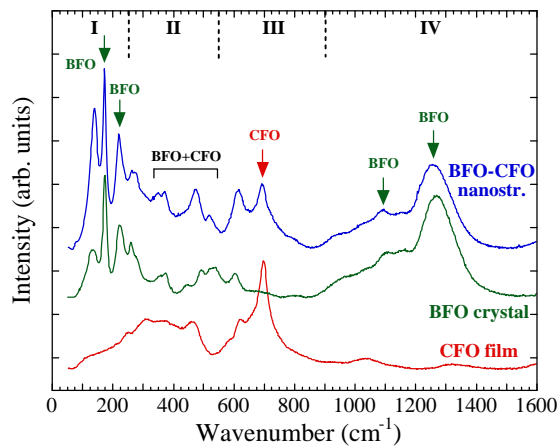

*Figure 1 (colour on-line)*

Comparison of 300 K Raman spectra for a 65BFO-35CFO nanostructure, a BFO single crystal and a CFO thin film. Regions I, II, III and IV denote specific spectral regions and the arrows indicate assigned Raman bands used for strain analysis.

In order to assign the Raman modes in the nanostructure, Figure 1 presents also reference Raman signatures taken from a BFO single crystal and a CFO thin film on STO (111). We parenthesize that BFO films on STO (111) are expected to adopt the bulk rhombohedral structure *R*3*c*, thus allowing us a meaningful comparison of the bulk and thin film Raman signatures. We first note that CFO presents only a very low intensity signature in the spectral regions I and IV or a signature of the STO substrate and, by simple comparison, the Raman bands of the nanostructure of these two regions can be directly attributed to a fingerprint of BFO, indicated by arrows in Figure 1. Note the comparable and remarkable sharpness of the BFO bands in the nanostructure and the bulk reference sample in region I, attesting the good crystalline quality of the BFO nano-pillars embedded in the nanostructure. Region II presents a more complex signature with overlapping and superimposing features of CFO and BFO, making a clear assignment unreliable. Finally, region III is characterized by two bands out of which the band at 694 cm$^{-1}$ can be assigned to CFO, while the band at 616 cm$^{-1}$ is a superposition of BFO and CFO components and thus difficult to analyse. The Raman spectrum of the nanostructure can thus be seen as a direct superposition of the Raman spectra of BFO and CFO. While this superposition complicates or even inhibits the interpretation of the spectral region II, the attributed features in regions I, III and IV (signed by arrows in Figure 1) allow a discussion of the strain state of both the BFO pillars and the CFO matrix, with respect to reference samples.

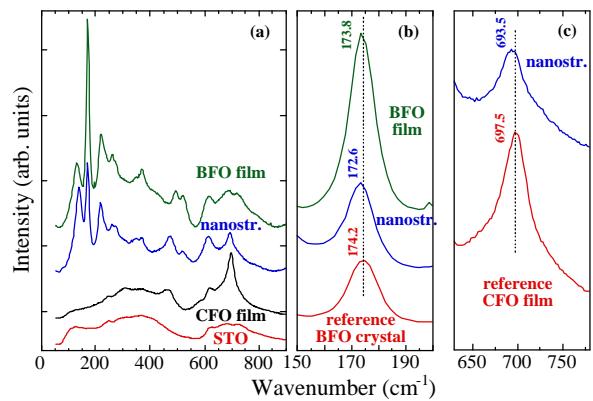

*Figure 2 (colour on-line)*

Raman spectra of a 65BFO-35CFO nanostructure, compared to simple thin films and the STO substrate (a) with a zoom on the BFO fingerprint band at ≈ 174 cm$^{-1}$ (b) and the CFO fingerprint band at ≈ 694 cm$^{-1}$ (c).

Strain manifests in a Raman spectrum mainly via shifts of the Raman bands. In absence of a close phase transition, an increase (decrease) in wavenumber is a sign of compressive (tensile) strain. In order to characterise the strain state of the pillar and matrix in the nanostructure, Figure 2.a compares a Raman spectrum of a 65BFO-35CFO thin film nanostructure to Raman spectra of BFO and CFO thin films on STO (111), respectively. The Figure shows also a spectrum of the STO (111) substrate, of which traces can be seen for the thin CFO film while they are negligible for the nanostructures. Figures 2.b-c present a detailed view of two specific bands around 174 and 694 cm$^{-1}$, which are compared to bulk and thin film spectra of BFO and CFO. The BFO crystal is expected to be strain-free, thus serving as a reference; similarly the thick CFO film, is used as reference for CFO in the nano-composite.

Figure 2.b shows small but distinctive difference in the position of the first order BFO Raman mode between the different samples, providing evidence for different strain states..

Both the nanostructure and the BFO thin film show a low wavenumber shift (thus tensile strain) with respect to the bulk reference.

The identification of the direction of the strain relies on the knowledge of the mode symmetry, an assignment which is chronically difficult in BiFeO$_3$,[23,24] with its close lying oblique modes of $A_1$ and $E$ symmetry at low wavenumber[24]. However, based on our observation of a strongly reduced intensity of low wavenumber mode at 174 cm$^{-1}$ under crossed polarisers (not shown), we assign an $A_1$ symmetry to this mode, thus a vibration along the polar rhombohedral direction (i.e. perpendicular to the thin film surface). This assignment is supported by the intensity ratio for oblique modes reported by Hlinka *et al.* on BFO bulk samples[24], allowing us to conclude that BiFeO$_3$ in the nanostructure present a tensile strain along the pillars axis with respect to the reference bulk sample.

Figure 2.c presents a comparison of the most intense CFO Raman mode. Under the common assumption of a *Fd-3m* cubic spinel structure, this strong mode is of $A_{1g}$ symmetry and can be assigned to Fe-O stretching vibrations along the cubic {111} direction of the FeO$_4$ tetrahedra (the so-called breathing mode) and is, as expected, the mode at the highest wavenumber.[15,26] For this mode, Iliev *et al.* [15] have proposed that the Fe-O bond lengths in NiFe$_2$O$_4$ scale with the length of the {111} space diagonals *d* and that the mode frequency will increase with decreasing *d* (and vice versa). Figure 2.c illustrates that the strain state of CFO in the pure CFO film is different from the CFO in the nanostructure as seen from the significant low wavenumber shift of the CFO phonon in the nanostructure. Based on the fact that the BFO pillars present an out-of-plane tensile strain, it is naturally expected that the CFO matrix presents an out-of-plane compressive strain, as we have verified by XRD. On the other hand, we observe a low wavenumber shift for the CFO band in the nanostructure, which lets us to suggest that the strong $A_{1g}$ at ≈ 694 cm$^{-1}$ of CFO presents a more complex relationship than that proposed in ref[15], which deserves further experimental and theoretical attention.

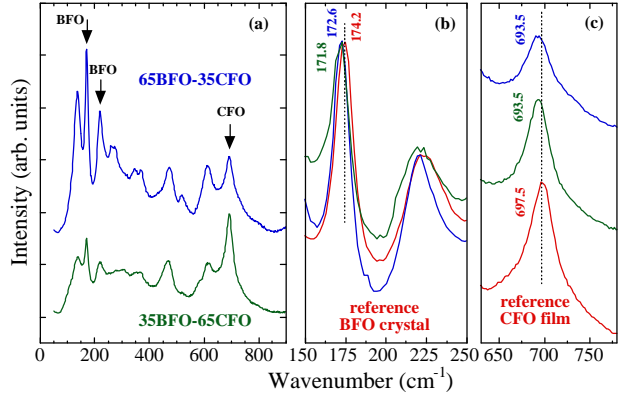

*Figure 3 (colour on-line)*

(a) Comparison of Raman spectra of 65BFO-35CFO and 35BFO-65CFO nanostructures. (b) BFO fingerprint region I, (c) CFO fingerprint region III. The spectra are normalized in (b) to facilitate the spectral comparison.

We now apply the same procedure to two nanostructures, which differ by the volume ratio between the perovskite and the spinel phase: a 65BFO-35CFO versus a 35BFO-65CFO nanostructure. At first sight the spectral signature of the two nanostructures in Figure 3.a is rather different, but this can be entirely explained by the different volume ratio, leading to a 65BFO–35CFO spectrum with strong BFO bands (regions I and IV) and a 35BFO–65CFO spectrum with a dominating CFO band at 694 cm$^{-1}$. Figures 3.b and 3.c compare specific bands of the two nanostructures to BFO and CFO reference data. Figure 3.b shows a zoom on the two low wavenumber $A_1$ symmetry BFO modes of the nanostructure, compared to the BFO bulk data. First, it is evident that the BFO pillars of both nanostructures undergo a low wavenumber shift when compared to the bulk, they are thus under tensile out-of-plane strain. Furthermore, a closer inspection by a spectral deconvolution shows that the BFO modes of the 35BFO–65CFO nanostructure are slightly more shifted and are also broader when compared to the 65BFO–35CFO nanostructure. The lower wavenumber can be understood by the fact that pillars are smaller and more isolated (rather than coalescent) in the 35BFO–65CFO nanostructure. As a consequence, it is to be expected that BFO pillars in the 35BFO–65CFO nanostructure are more sensitive to the CFO presence than in the latter, thus leading naturally to an increased strain. The slight broadening might be related either to a shorter coherence length of the isolated pillars or to the presence of different strain states in pillars of different thickness. On the other hand, the CFO matrix presents the same strain state in both nanostructures as deduced from the same position of the CFO $A_{1g}$ mode.

In summary, we have presented a Raman scattering investigation of BFO-CFO thin film nanostructures. We have shown that in spite of the inherent complexity of the resulting

Raman spectra, distinctive strain states can be identified depending on the film composition. Due to the extreme versatility of Raman spectroscopy, we envisage that experiment aim to monitor electric/magnetic-induced changes of strain states and thus elastic coupling could be possible. Our experimental investigation demonstrates that Raman scattering can be used not only for the analysis of multiferroic multilayers but also for three dimensional multiferroic pillar/matrix composites and we expect that it can be extended to other complex multiferroic composites such as the recently reported core-shell nanoparticles and nanotubes.[27]

*Acknowledgements*

P. Jadhav acknowledges an Erasmus Mundos Postdoctoral Fellowship within the External Cooperation Window program. Financial support by the Ministerio de Ciencia e Innovación of the Spanish Government [Projects MAT2008-06761-C03 and NANOSELECT CSD2007-00041] and Generalitat de Catalunya (2009 SGR 00376) is acknowledged.